\documentclass[manuscript]{aastex}
\usepackage{CJK}
\usepackage{times}
\usepackage{helvet}
\usepackage{courier}
\usepackage{amssymb}
\usepackage{amsmath}
\usepackage{multirow}
\usepackage{amsopn}
\usepackage{epsfig}
\usepackage{graphicx,subfigure}
\usepackage{array}
\usepackage{url}
\usepackage{bm}
\usepackage{algorithm}
\usepackage{tabularx}

\begin{document}

\title{An Efficient Method for Rare Spectra Retrieval in Astronomical Databases}

\author{Changde Du\altaffilmark{1,2}}
\affil{School of Physics, University of Chinese Academy of Sciences, Beijing 100049, China}
\author{Ali Luo\altaffilmark{1,2}}
\affil{Key Laboratory of Optical Astronomy, National Astronomical Observatories, Chinese Academy of Sciences, Beijing 100012, China}
\email{lal@nao.cas.cn}
\author{Haifeng Yang\altaffilmark{1,3}, Wen Hou\altaffilmark{1,2}, and Yanxin Guo\altaffilmark{1,2}}
\affil{University of Chinese Academy of Sciences, Beijing 100049, China}

\altaffiltext{1}{Key Laboratry of Optical Astronomiy, National Astronomical Observatories, Chinese Academy of Sciences, Beijing 100012, China.}
\altaffiltext{2}{University of Chinese Academy of Sciences, Beijing 100049, China.}
\altaffiltext{3}{School of Computer Science and Technology, Taiyuan University of Science and Technology, Taiyuan 030024, China.}

\begin{abstract}
One of important aims of astronomical data mining is to systematically search for specific rare objects in a massive spectral dataset, given a small fraction of identified samples with the same type.  Most existing methods are mainly based on binary classification, which usually suffer from  uncompleteness when the known samples are too few.  While, rank-based methods would provide good solutions for such case. After investigating several algorithms, a method combining bipartite ranking model with bootstrap aggregating techniques was developed in this paper.  The method was applied in searching for carbon stars in the spectral data of Sloan Digital Sky Survey (SDSS) DR10, and compared with several other popular methods used in data mining. Experimental results validate that the proposed method is not only the most effective but also less time consuming among these competitors automatically searching for rare spectra in a large but unlabelled dataset.

\end{abstract}


\section{Introduction}

In many applications, only a finite set of interesting samples sharing common traits are given, and the goal is to search other samples sharing the same traits from datasets. For example, in rare astrophysical objects searching, the finite set can be a series of spectra  known as a specific unusual class comparing with main sequence stars(such as carbon stars, DZ white dwarfs, L dwarfs, etc.), and the goal is to search for as many as spectra with the same class in massive astronomical data sets. In this application, the number of positive (interesting but rare) samples are limited, while the unlabelled samples dominate the dataset.

Conceptually, learning from the positive and unlabeled samples is usually called PU (positive-unlabeled) learning which arises frequently in retrieval applications. Formally, we assume that $\mathbf{X}=\{x_{1},\ldots, x_{p+u}\}$ denotes a set of data belonging to the instance space $\mathcal{X}=\{x \in \mathbb{R}^d \}$, $\mathbf{P}=\{x_1,\ldots, x_p \}$ is a set of (usually small) data with positive labels, and $\mathbf{U}=\{x_{p+1},\ldots, x_{p+u}\}$ is the set of (usually large) unlabeled data. Then we need to learn from $\mathbf{P}$ and $\mathbf{U}$ in order to identify the positive samples from $\mathbf{U}$ as accurate as possible. The purpose of PU learning is to learn a scoring function $f: \mathcal{X} \rightarrow \mathbb{R}$ from $\mathbf{P}$ and $\mathbf{U}$,  which is able to predict a score to each unlabeled data in $\mathbf{U}$. For $\forall$  $x_i\in\mathbf{U}$, the higher the prediction score $f(x_i)$ indicates  the larger probability to be a positive sample.

There are many different algorithms developed to solve the PU learning problem over the last decade, and we summarise them into two groups: classification-based PU learning and rank-based PU learning.

The classification-based PU learning could be traced back to building classifier with only positive sample, such as One-Class SVM (OCSVM) \citep{scholkopf1999support, manevitz2002one} and Support Vector Data Description (SVDD) \citep{tax2004support}.  Both OCSVM and SVDD need sufficient positive samples to garentee the boundary of the positive class can be induced precisely. Aside from the positive samples, unlabeled samples can also provide useful information, and have been used along with the positive samples by Biased SVM \citep{liu2003building}.   Mordelet et al. \citep{mordelet2014bagging} generalised Biased SVM  and proposed a method employing bootstrap aggregating (bagging) techniques \citep{breiman1996bagging}, called Bagging SVM. The author showed that Bagging SVM method can match and even outperform the performance of Biased SVM. Furthermore, Bagging SVM can greatly alleviate the computation burden of Biased SVM, in particular when unlabeled samples domainate the dataset. To the best of our knowledge, Bagging SVM represents the state-of-the-art algorithm for PU learning.

For rank-based PU learning, the core idea is to build a ranking model which ranks a set of unlabeled samples based on their relevance scores for the given positive samples. Particularly, the graph-based ranking method is widely used in PU learning, such as Label Propagation (LP) \citep{zhou2004learning} and Manifold Ranking (MR) \citep{zhou2004ranking}. These methods have been applied to search for carbon stars from massive spectral data, successfully retrieving 260  and 183 carbon stars from SDSS DR8 and Large Sky Area Multi-Object Fiber Spectroscopic Telescope (LAMOST), respectively \citep{si2014search,Si2015carbon}. Other recent works have considered the PU learning as the bipartite ranking problem \citep{amini2008boosting,61acebb7c98a4881b4c72e29e6d6ea8d,kotlowski2011bipartite}. Specifically, the negative samples are chosen from $\mathbf{U}$ according to some rules, such as similarity measure \citep{amini2008boosting} and random sampling \citep{61acebb7c98a4881b4c72e29e6d6ea8d}. Once a sample in $\mathbf{U}$ is chosen as a negative one, it will be assigned a negative label in the training stage. After generating procedure for negative samples, the positive and negative samples in the rest of $\mathbf{U}$ are called relevant and irrelevant samples, respectively. Then, they develop models ranking positive samples ahead of the chosen negative samples, based on the assumption that such models would also place the other relevant samples ahead of irrelevant samples.

In this paper, we treat the rare spectra retrieval as the bipartite ranking task, and present a new PU learning method to solve this problem. More precisely, inspired by the idea of bagging techniques which have shown to be useful for improving the stability and accuracy of machine learning algorithms \citep{mordelet2014bagging}, we developed a new method, namely, BaggingTopPush that combines bagging techniques with TopPush model \citep{NIPS2014_5222}. It is worth noting that BaggingTopPush method focus on optimising ranking accuracy at the top of the ranked list. This fulfils the scientific need for an accurately top ranked list of candidates under a given rare category. Furthermore, BaggingTopPush is a highly efficient approach that has computational complexity linear in the number of training samples. In rare spectra retrieval, we usually only have a small set of positive samples and a large set of unlabeled samples. In other words, it does not contain any explicit set of negative samples (irrelevant spectra), and it is time-consuming to manually select negative ones frequently. Besides that, a small portion of the negative samples selected manually may not be able to represent the overall nature of the negative data. Here, we follow the method used by Mordelet et al. \citep{mordelet2011prodige} and randomly select the unlabeled spectra to generate `negative' samples. Then the BaggingTopPush method consists in aggregating bipartite ranking functions trained to place $\mathbf{P}$ before a small random subsample of $\mathbf{U}$. In order to check the effectiveness and efficiency of the method for rare spectra retrieval, several popular PU learning methods are introduced and compared with the proposed one in this paper. For easier application, the influence of BaggingTopPush's parameters on the ranking performance were also analysed to obtain safe parameter choices.

The paper is organised as follows. In section 2, the TopPush method based on bipartite ranking is briefly described firstly, and then the development of a bagging strategy for rare spectra retrieval is presented.  In section 3, the experimental data, comparative methods, parameters setting, and evaluation metrics are given.  In section 4, the detailed experimental evaluation and comparison of several PU learning methods are shown. Finally, the conclusion is drawn in section 5.

\section{Bipartite ranking model}
 In recent years, bipartite ranking has attracted much attention because of its successful  applications in several areas such as information retrieval and recommendation systems \citep{liu2009learning,rendle2009learning}. The goal of bipartite ranking is to learn a ranking model such that those samples belonging to one category are ranked higher than those samples belonging to the other category. In some data mining applications such as web page searching and rare spectra retrieval, to learn a ranking function with well performance at the top of the ranked list are more interested since only the top ranked candidates are possible to be examined by experts\citep{clemenccon2007ranking,boyd2012accuracy}.

 TopPush proposed by Li et al. \citep{NIPS2014_5222} is such a bipartite ranking model that can efficiently optimise the ranking accuracy at the top. In contrast to most other methods for bipartite ranking whose computational costs grow quadratically in the number of training samples, the time complexity of TopPush algorithm is only linear in the number of training samples. In the following of this section, we will first describe the TopPush algorithm, and then develop the bagging strategy used to retrieve rare spectra.

\subsection{TopPush method}
Let $\mathbf{S}=\mathbf{S}_{+} \cup \mathbf{S}_{-}$ be a set of training data, including $m$ positive samples and $n$ negative samples randomly sampled from $\mathbf{P}$ and $\mathbf{U}$, respectively, i.e., $\mathbf{S}_{+}=\{x^{+}_i\in P\}_{i=1}^m$  and $\mathbf{S}_{-}=\{x^{-}_j\in U\}_{j=1}^n$. The goal of TopPush is to learn a ranking function $f: \mathcal{X} \rightarrow \mathbb{R}$ that is likely to maximise the number of positive samples that are ranked before the first negative sample. This objective can be translated into the minimisation of the following loss function
\begin{equation} \label{TopPush loss}
\begin{array}{@{ }l@{\quad}l}
\begin{aligned}
\mathcal{L}(f;\mathbf{S})=\frac{1}{m}\sum\limits_{i=1}^{m} \mathbb{I}(f(x^{+}_i)\leq \max \limits_{1\leq j \leq n} f(x^{-}_j)),
\end{aligned}
\end{array}
\end{equation}
where $\mathbb{I}(\cdot)$ is the indicator function with $\mathbb{I}(true) = 1$ and 0 otherwise.
By minimising the loss function in equation (\ref{TopPush loss}), it essentially pushes negative samples away from the top of the ranked list, leading to more positive ones placed at the top. Since the indicator function $\mathbb{I}(\cdot)$ is not a smooth function, Li et al. replaced the indicator function in equation (\ref{TopPush loss}) with its convex surrogate loss function $\ell(\cdot)$ that is non-decreasing and differentiable, leading to the following loss function
\begin{equation} \label{TopPush surrogate loss}
\begin{array}{@{ }l@{\quad}l}
\begin{aligned}
\mathcal{L}(f;\mathbf{S})=\frac{1}{m}\sum\limits_{i=1}^{m} \ell(\max \limits_{1\leq j \leq n} f(x^{-}_j) - f(x^{+}_i)).
\end{aligned}
\end{array}
\end{equation}
In practice, the convex surrogate loss functions include truncated quadratic loss $\ell(z)= \max(0,1+z)^2$, exponential loss $\ell(z)=  e^z$ and logistic loss $\ell(z)=\mathrm{log}(1 + e^z)$, etc. Here, we restrict ourselves to the truncated quadratic loss function.

For linear ranking function ($f(x) = \mathrm{\mathbf{w}}^T x$), the learning problem is given by the following optimisation formulation
\begin{equation} \label{optimizasion problem}
\begin{array}{@{ }l@{\quad}l}
\begin{aligned}
\min\limits_{\mathrm{\mathbf{w}}}\   \frac{\lambda}{2}\|\mathrm{\mathbf{w}}\|^2 +\frac{1}{m}\sum\limits_{i=1}^{m} \ell(\max \limits_{1\leq j \leq n} \mathrm{\mathbf{w}}^Tx^{-}_j - \mathrm{\mathbf{w}}^Tx^{+}_i),
\end{aligned}
\end{array}
\end{equation}
where $\mathrm{\mathbf{w}}\in \mathbb{R}^d$ is the weight vector to be learned, and $\lambda>0$ is a regularisation parameter that controls the model complexity. More discussions about the optimisation, computational complexity and performance guarantee for TopPush algorithm can be found in \citep{NIPS2014_5222}.
\subsection{Bagging TopPush for rare spectra retrieval}
In rare spectra retrieval, given some rare spectra our final objective is to rank relevant samples ahead of irrelevant samples. To this end, one key assumption is that learning to place $\mathbf{P}$ (rare spectra) before a small random subsample of $\mathbf{U}$ (unlabeled spectra) is a good proxy to our objective. However, the unlabeled spectra set $\mathbf{U}$ is contaminated by hidden positive spectra, and the percentage of positive spectra in $\mathbf{U}$ is usually unknown in real-world applications.  For a small random subsample of $\mathbf{U}$, the contamination (percentage of rare spectra) can be small or large, which will induce a large instability in the ranking function. Fortunately, this situation can be advantageously exploited by bagging techniques which are designed to improve the stability and accuracy of unstable machine learning algorithms \citep{breiman1996bagging}.

We assume that $K$ is the number of samples randomly selected from $\mathbf{U}$ by one bootstrap and $T$ is the number of bootstraps. The BaggingTopPush first creates a series of  bipartite ranking function trained to rank $\mathbf{P}$ ahead of the random subsamples of $\mathbf{U}$. The output of each of these bipartite ranking functions $f_t$ can assign a ranking score to any sample. Then the final aggregated ranking function $f$ can be simply defined as the average score of the individual bipartite ranking functions, and we can sort spectral samples according to $f$ in a descending order and return the top ranked ones as results. In summary, the BaggingTopPush method for rare spectra retrieval is presented in Algorithm 1. Note that the input variable $\lambda$ plays the same role as that in equation (\ref{optimizasion problem}), i.e., it controls the complexity of each TopPush model. The smaller the value of $\lambda$, the more complicated model we have, i.e., the more time will be consumed at the training stage.
\begin{algorithm} [ht] \footnotesize
\caption{BaggingTopPush for rare spectra retrieval}
\textbf{Input:} $\mathbf{P}$, $\mathbf{U}$, $K$, $T$, $\lambda$.\\
\textbf{Output:} Ranking function $f: \mathcal{X} \rightarrow \mathbb{R}$.
\begin{enumerate}
   \item \textbf{for} $t=1$ to $T$ \textbf{do}
           \begin{itemize}
                \item Draw a subsample $U_t$ of size $K$ from the set of unlabeled spectra $\mathbf{U}$.
                \item Train a TopPush model $f_t$ to rank $\mathbf{P}$ ahead of $U_t$ .
           \end{itemize}
%   \item \textbf{end for}
   \item \textbf{return} $f=\frac{1}{T}\sum\limits_{t=1}^T f_t$
\end{enumerate}
\end{algorithm}
\section{Experiments}
In this section, we conducted a set of rare spectra retrieval experiments and presented detailed experimental comparison to evaluate the effectiveness and efficiency of various methods. Specifically, we investigated the influence of different model parameter choices for our BaggingTopPush method to make a reference for simply application.

\subsection{Experimental data}
We use the spectral data set taken from the Sloan Digital Sky Survey (SDSS)\footnote{http://www.sdss3.org/dr10/}, Data Release 10 (DR10) \citep{ahn2014tenth}, and choose the carbon stars as the rare astrophysical objects to demonstrate the performance of our method. Note that other types of rare spectra can also be retrieved by our method, and here we focused on carbon stars. Specifically, we carefully select 450 samples from all carbon stars classified by the spectroscopic pipeline of SDSS, and randomly select 100,000 samples from other types of stellar spectra. Since carbon stars are quite rare, we roughly consider that there are no carbon stars mixed in these 100,000 stellar spectra. All spectral data have the wavelength coverage from 3917.4 \AA\ to 8974.3 \AA\ and thus the dimensionality is 3601.

\begin{figure}[!htbp]
\centerline{\includegraphics[height=3.0in,width=5in]{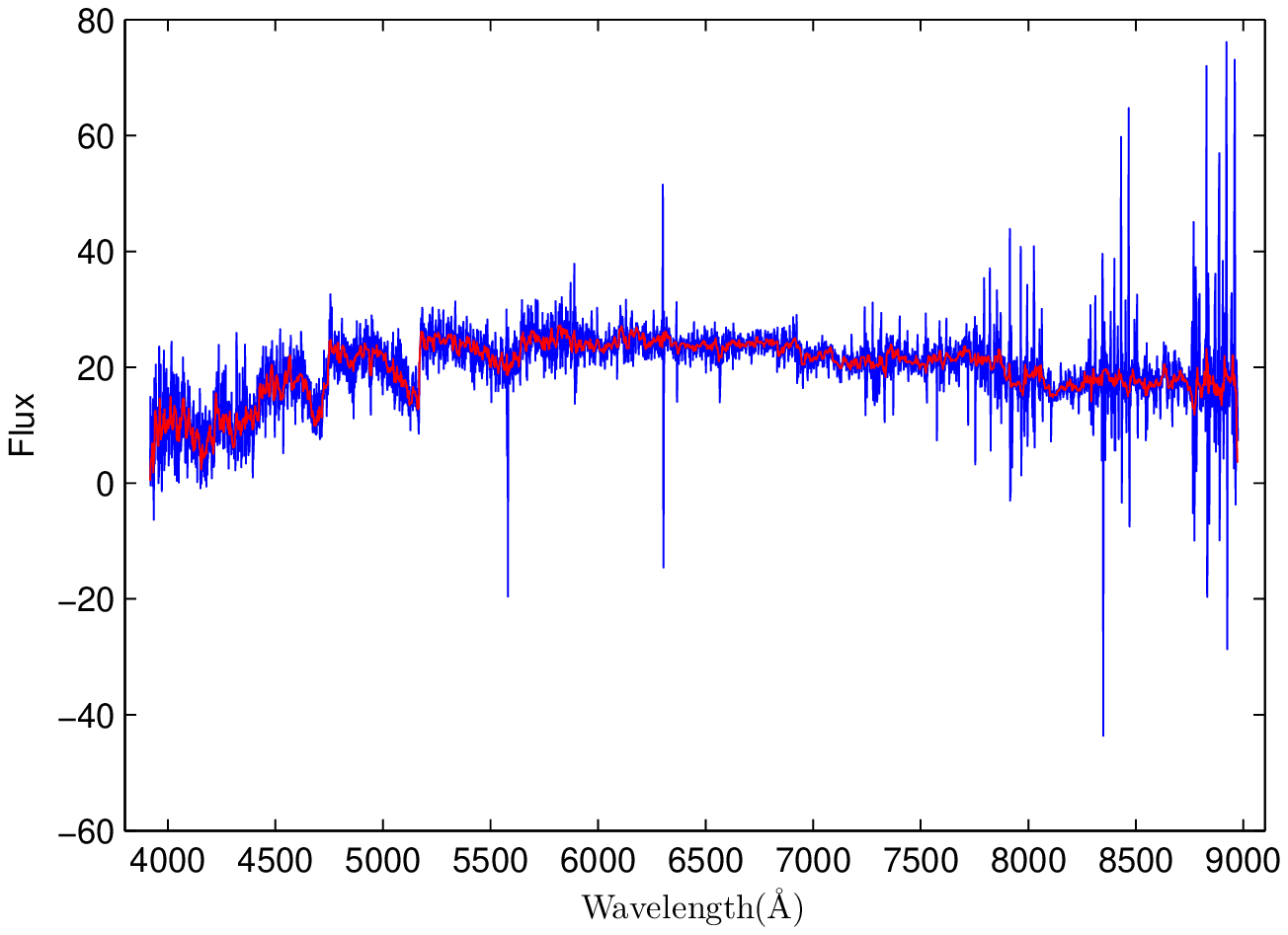}}
\caption{To deal with the spectra by a median filter with width 10 \AA. The blue line is original spectral data and the red line is filtered spectral data.}
\label{denoising}
\end{figure}

Generally, the spectral data should be preprocessed to facilitate retrieval. The preprocessing includes de-noising, normalisation, feature selection, etc. In the experiments, we first employ a median filter with width 10 \AA\ to eliminate the disturbance of narrow skylines and noise. One filtered example is  presented in Fig. \ref{denoising}. Obviously, the skylines and noise have been removed effectively after filtering.  We then normalise the spectral flux by mapping the minimum and maximum flux value of each spectral data to $[0, 1]$. This is very useful for data mining applications where the input data are generally distributed on widely different scales. In rare astrophysical objects searching, we are often confronted with very high dimensional spectral data which may contains a lot of non-informative or noisy features, so it is necessary to extract the main information hidden in the spectral data. We finally apply the Principal Component Analysis (PCA) which has been widely used in spectra classification problems to obtain the low-dimensional data representation, and 50 principal components have been retained.

\subsection{Comparative methods}
In this subsection, we briefly introduce some other PU learning methods which can be used to retrieve rare spectra. We select One-Class SVM (OCSVM) and Bagging SVM (BaggingSVM) to serve as the classification-based methods, in which BaggingSVM represents the state-of-the-art one. Furthermore, Label Propagation (LP), Efficient Manifold Ranking (EMR) and Local Regression and Global Alignment (LRGA) can serve as the rank-based methods.

The formulation of OCSVM proposed by Sch$\mathrm{\ddot{o}}$lkopf et al. can be summarized as mapping the data (only positive class) into a feature space $H$ using an appropriate kernel function, and then trying to find a hyperplane to separate the mapped vectors from the origin with maximum margin \citep{scholkopf1999support,manevitz2002one}.  Then for a new, previously unseen sample, its label can be determined by  this hyperplane.

Before we introduce the Bagging SVM, let us briefly cover the Biased SVM \citep{liu2003building}. Biased SVM treats all the unlabeled samples as negative samples. Then the SVM classifier is built by giving appropriate weights to the positive and unlabeled samples, respectively. Based on the idea of bagging and Biased SVM, Mordelet et al. \citep{mordelet2014bagging} proposed Bagging SVM method, which consists in aggregating Biased SVM classifiers trained to discriminate $\mathbf{P}$ from a small random subsample of $\mathbf{U}$. The motivation behind bagging SVM is to exploit an intrinsic feature of PU learning to benefit from classifier aggregation through a random subsample strategy, and we borrowed this idea in our BaggingTopPush method.

LP proposed by Zhou et al. \citep{zhou2004learning} is one of the state-of-the-art graph-based learning algorithms. Si et al. \citep{si2014search} have successfully applied the LP algorithm to search carbon stars and DZ white dwarfs from Data Release 8 (DR8) of the Sloan Digital Sky Survey (SDSS). Specifically, they have found 260 new carbon stars and 29 new DZ white dwarfs from SDSS DR8. The key assumption in LP is that neighboring data points in high dimensional space should share the similar semantic labels. Given a set of data, the generated graph is often represented as an adjacency matrix in which the elements save the edge weights between any two points. Generally, the k-NN graph scheme is the most popular approach for graph construction. In the k-NN graph, the weights are usually defined by the Gaussian kernel.

EMR proposed by Xu et al. \citep{xu2011efficient} is a new framework for large scale retrieval problems. Si et al. \citep{Si2015carbon}  have applied the EMR to search carbon stars from the Large Sky Area Multi-Object Fiber Spectroscopic Telescope (LAMOST) pilot survey. Using this algorithm, they totally found 183 carbon stars, and 158 of them are new findings.
The goal of EMR is to address the shortcomings of MR \citep{zhou2004ranking} from scalable graph construction and efficient computation. Specifically, EMR precomputes an anchor graph on the data set instead of the traditional k-NN graph to enhance the computation speed of MR.

The normalized Laplacian matrix in LP is usually calculated based on Gaussian function. However, it has been reported that Gaussian function is sensitive to the width parameter \citep{wang2008label}, and there is usually no truth to tune the width parameter of Gaussian kernel in real-world retrieval applications. To overcome this problem, LRGA \citep{yang2009ranking} learns the Laplacian matrix by local regression and global alignment and shows to be insensitive to parameters.

\subsection{Experimental settings}
To simulate a PU learning problem, we randomly select a given number of carbon star spectra to create a positive set $\mathbf{P}$, while $\mathbf{U}$ contains the non-selected carbon star spectra and all of the other spectra. To investigate the influence of the number of known positive samples, we varied the size of $\mathbf{P}$ (NP) in $\{1, 3, 5, 10, 15, 30, 50\}$. For each value of NP, we trained all 6 methods described above (BaggingTopPush,  OCSVM, BaggingSVM, LP, EMR, LRGA) and ranked the spectra in $\mathbf{U}$ by decreasing score. The parameters of all 6 algorithms were carefully tuned on our data. For BaggingTopPush, we varied the regularisation parameter $\lambda$ that controls the model complexity in $\{0.001, 0.01, 0.1, 1, 10, 100, 1000\}$, the size of bootstrap samples $K$ in $\{1, 5, 10, 20, 50, 100, 200, 300\}$ and the number of bootstraps $T$ in $\{1, 5, 10, 15, 25, 40, 60, 80, 100\}$. For OCSVM, the parameter $\nu$ denotes an upper bound on the fraction of outliers (training examples regarded out-of-class) and a lower bound on the fraction of training examples used as support vectors, and we varied $\nu$ in $\{0.1, 0.2, 0.3, 0.4, 0.5\}$. For BaggingSVM, the penalty parameter $C$ which determines the influence of the misclassification on the objective function was chosen from $\{e^{-8}, e^{-6}, \ldots, e^{8}\}$, the size of bootstrap samples equals to NP, and the number of bootstraps equals to 30. For LP and LRGA, we constructed the k-NN graph where k$=$5. Furthermore, in LP the width parameter of Gaussian kernel was set to $\sigma=1$, and the weight parameter $\alpha$ which balances the smoothness constraint and the fitting constraint in objective function was set to 0.99. For EMR, we set the number of anchors to $d=1000$, and the weight parameter $\alpha$ to 0.99 as well, which consistent with the experiments performed in \citep{xu2011efficient}.

All methods are implemented in MATLAB environment on a workstation with 12-core Intel(R) Xeon(R) (3.47GHz) with 96 GB RAM. The implementations of OCSVM and BaggingSVM are based on the LIBSVM\footnote{http://www.csie.ntu.edu.tw/\%7Ecjlin/libsvm/.} package \citep{CC01a}. We perform 50 independent trials for each algorithm, and the averaged results are reported.
\subsection{Evaluation metrics}
Each spectral sample in the testing set is ranked based on its prediction score. Due to time constraint, only a small set of the top ranked spectra will be validated by astronomer.  It indicates that a best ranking means all the relevant spectra should be ranked in the top positions. Hence, we evaluate the models performance using the following several information retrieval metrics, which mainly focus on behaviour at the top of a ranked list.
\begin{itemize}
\item The precision at $k$ ($\mathrm{P}@k$): It measures what fraction of the top $k$ ranked spectra belong to the given rare category.
\item The recall at $k$ ($\mathrm{R}@k$): It measures what fraction of the known rare spectra are retrieved within the top $k$ ranked spectra.
\item The average precision at $k$ ($\mathrm{AP}@k$): by computing the precision and recall at every position in the ranked list of spectra, one can plot a precision-recall (PR) curve, and AP denotes the area under the PR curve. $\mathrm{AP}@k$ is defined as:
\begin{equation} \label{AP@k} \nonumber
\begin{array}{@{ }l@{\quad}l}
%\begin{aligned}
\mathrm{AP}@k=\frac{\sum_{l=1}^k\mathbb{I}(x_l \in \mathbf{U}_r)\mathrm{P}@l}{\mathrm{min}(|\mathbf{U}_r|,k)},
%\end{aligned}
\end{array}
\end{equation}
where $\mathbf{U}_r$ denotes the set of relevant spectra in unlabelled data set $\mathbf{U}$ and $|\mathbf{U}_r|$ is the size of $\mathbf{U}_r$.
\item The area under the receiver operating characteristic curve (AUC): It measures the global ranking performance of the model, wherever the incorrect pair-wise ordering occurs in the ranking list.
\end{itemize}
\section{Results}
\subsection{Ranking effectiveness}
\begin{figure*}[!htbp]
\setlength{\abovecaptionskip}{0pt}
\setlength{\belowcaptionskip}{-10pt}
\centering
\subfigure[NP=1] {\includegraphics[height=1.7in,width=2.1in]{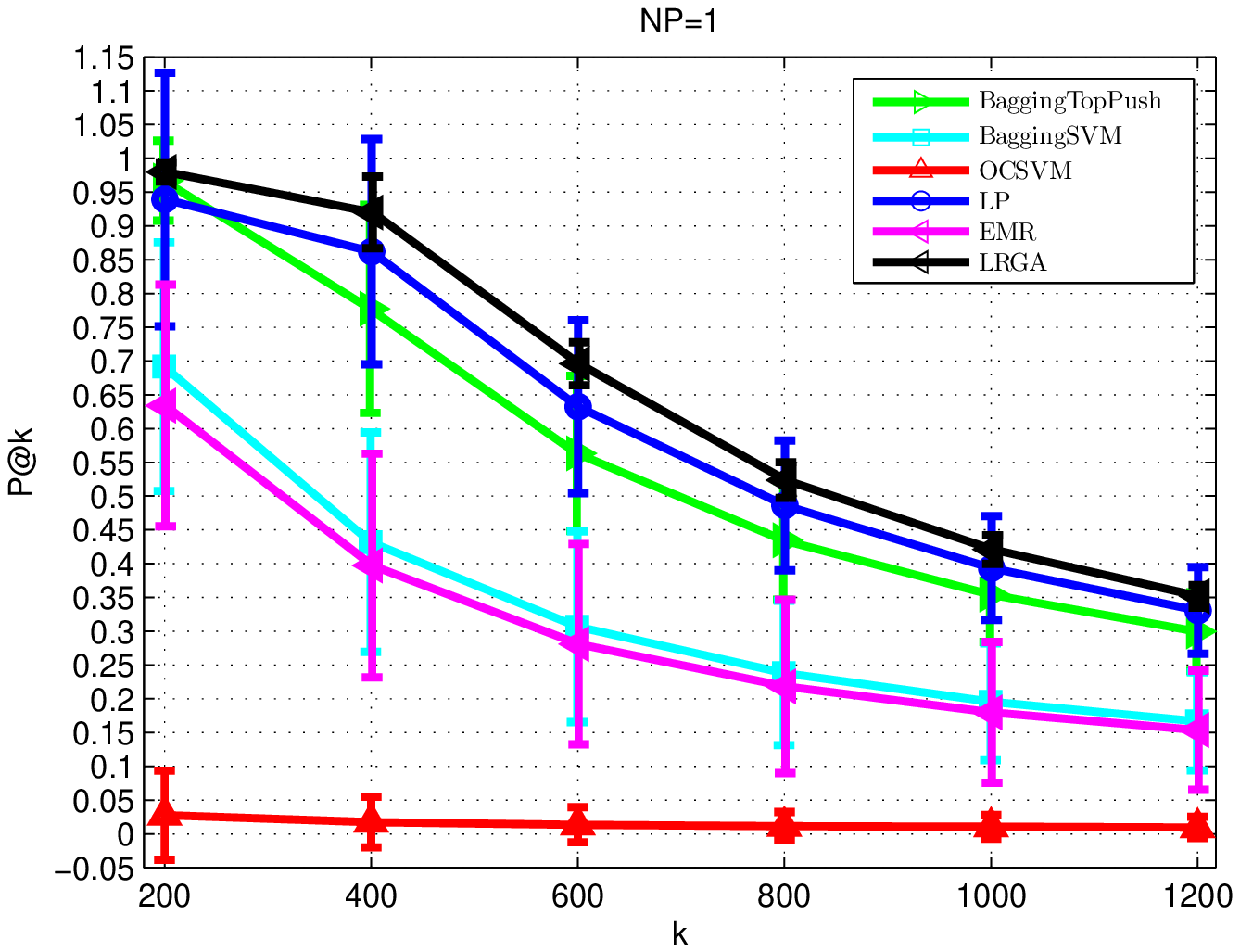}}
\subfigure[NP=10] {\includegraphics[height=1.7in,width=2.1in]{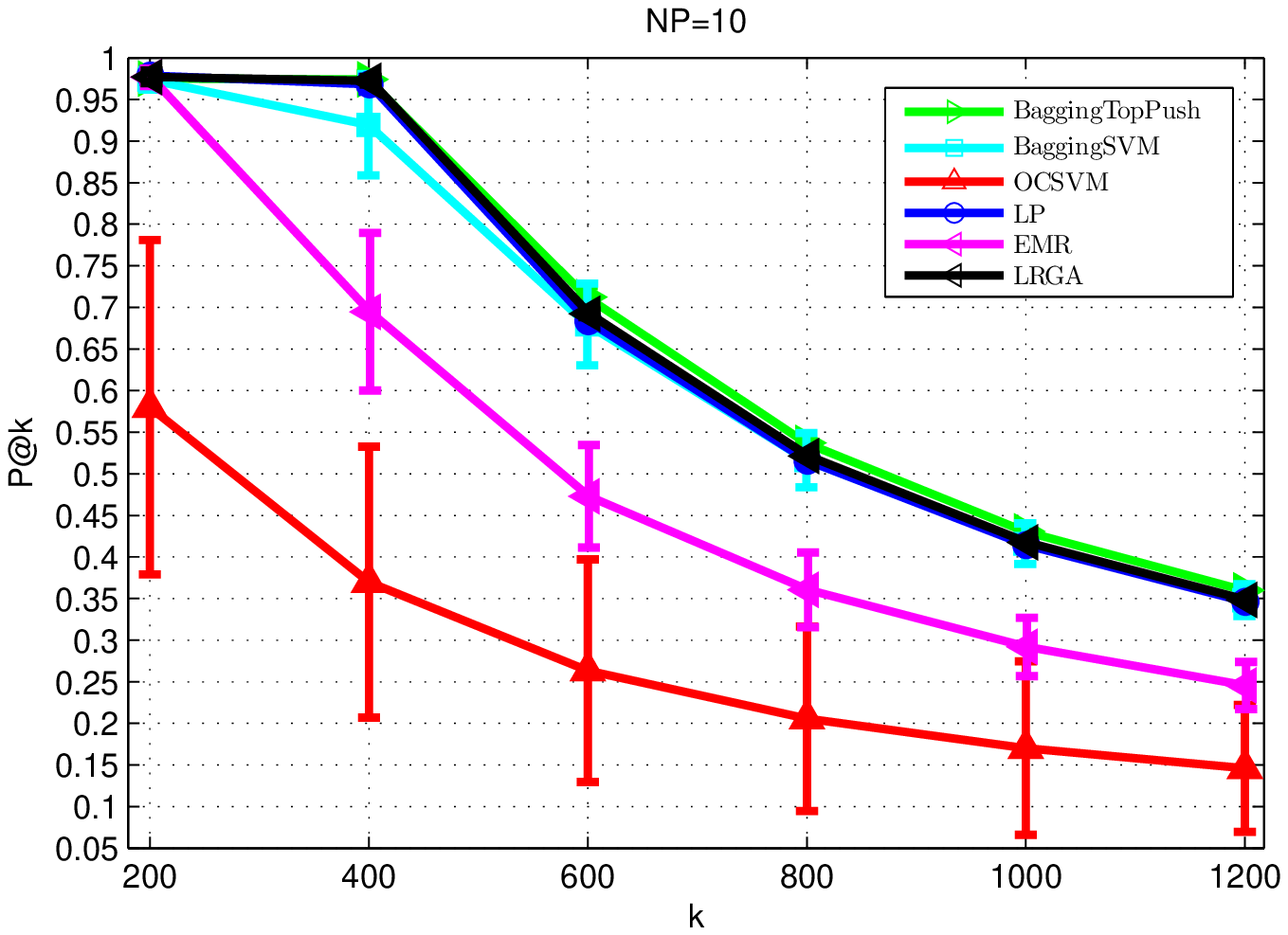}}
\subfigure[NP=30] {\includegraphics[height=1.7in,width=2.1in]{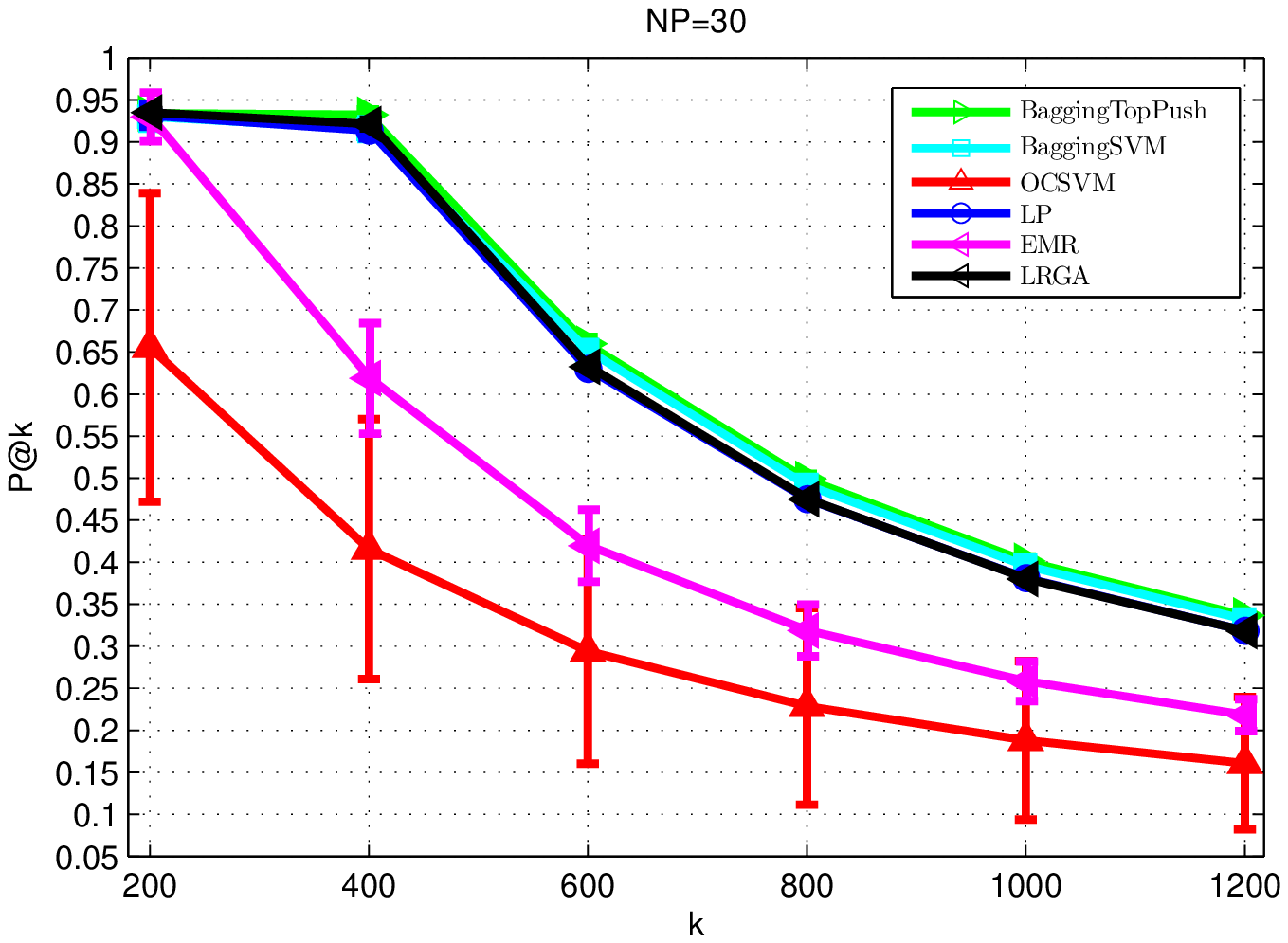}}
\caption{A comparison of average $\mathrm{P}@k$ for different settings of $k$ and the number of positive training samples NP.}
\label{fig2}
\end{figure*}

\begin{figure*}[!htbp]
\centering
\subfigure[NP=1] {\includegraphics[height=1.7in,width=2.1in]{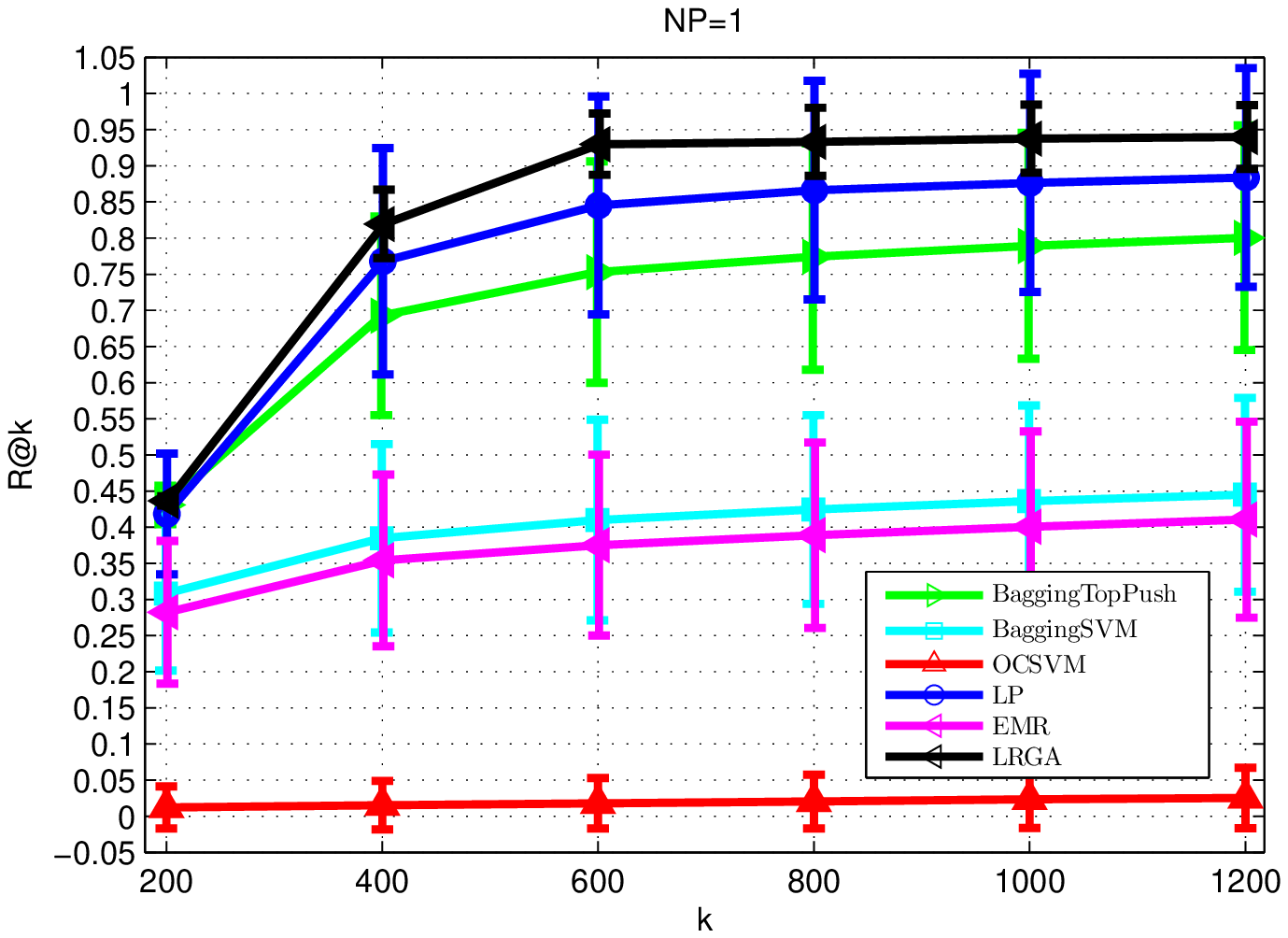}}
\subfigure[NP=10] {\includegraphics[height=1.7in,width=2.1in]{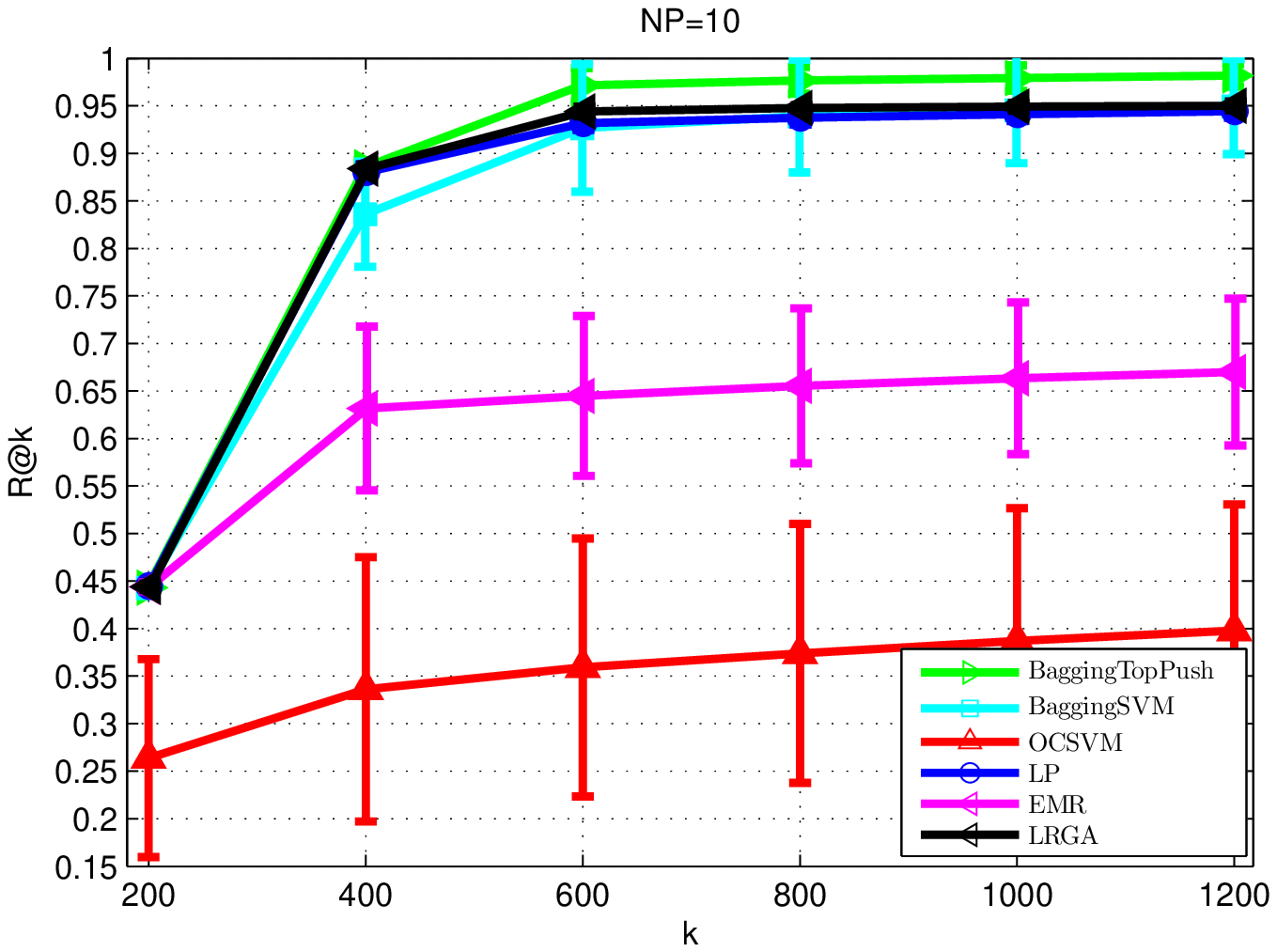}}
\subfigure[NP=30] {\includegraphics[height=1.7in,width=2.1in]{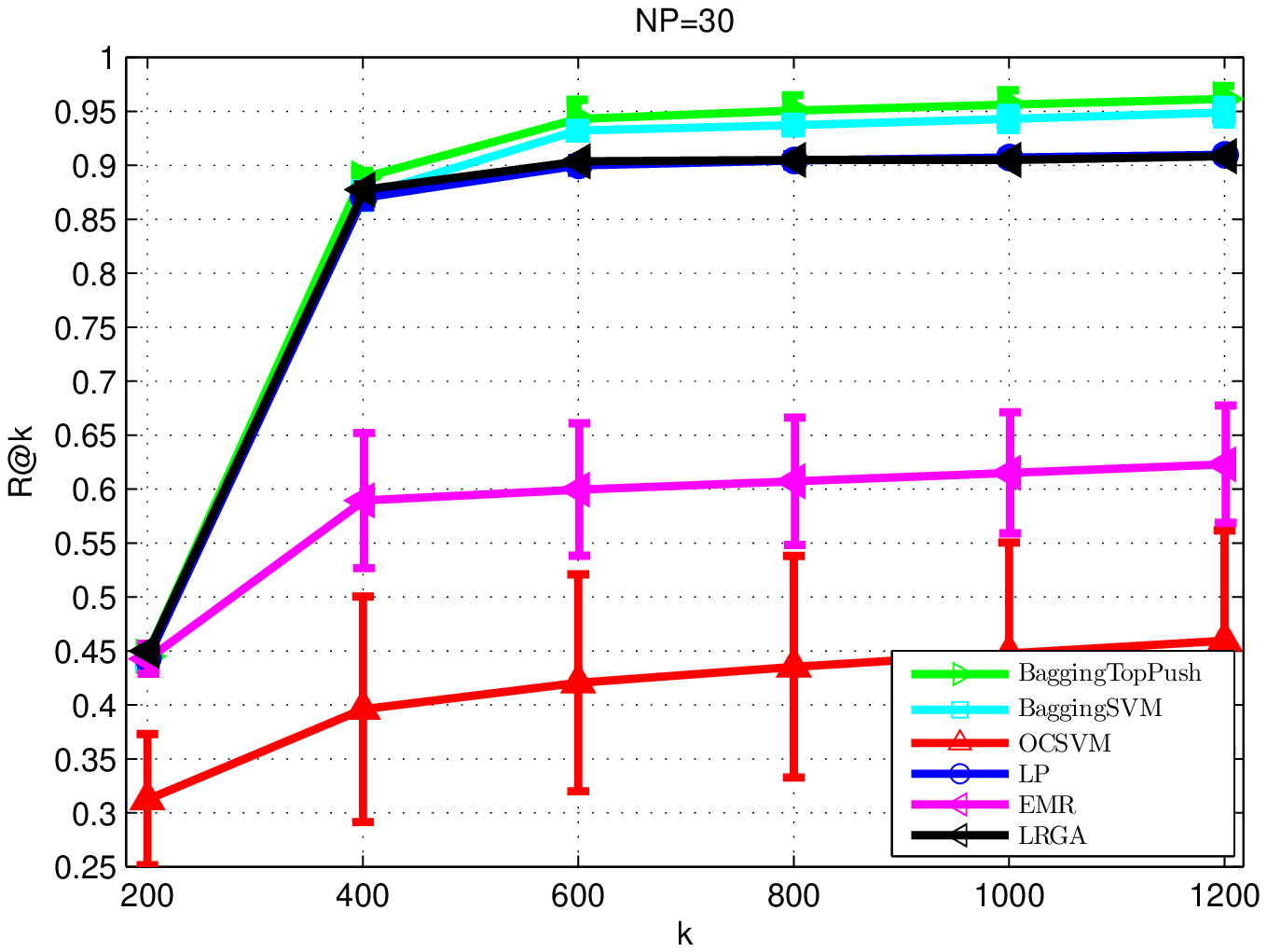}}
\caption{A comparison of average $\mathrm{R}@k$ for different settings of $k$ and the number of positive training samples NP.}
\label{fig3}
\end{figure*}

\begin{figure*}[!htbp]
\centering
\subfigure[NP=1] {\includegraphics[height=1.7in,width=2.1in]{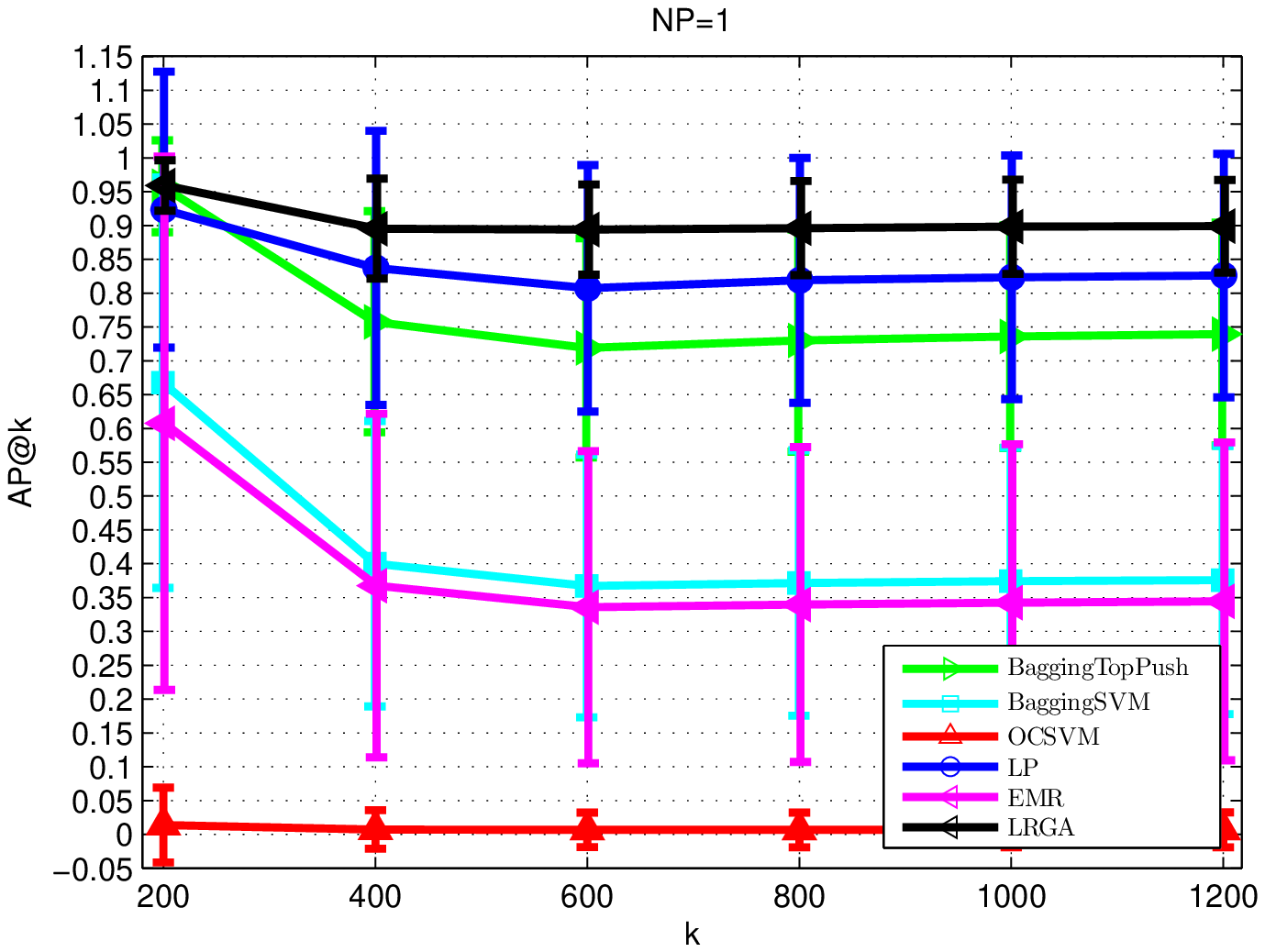}}
\subfigure[NP=10] {\includegraphics[height=1.7in,width=2.1in]{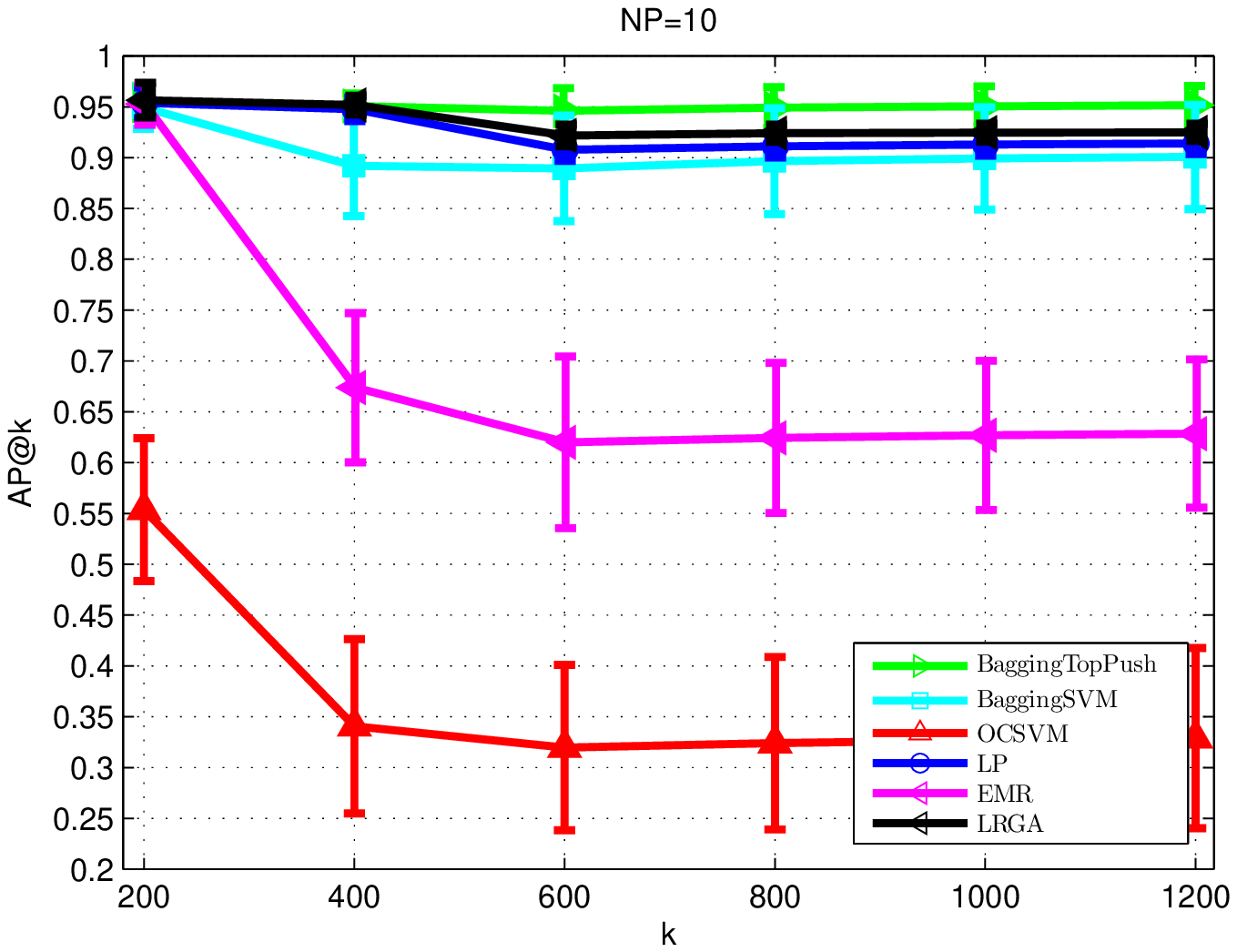}}
\subfigure[NP=30] {\includegraphics[height=1.7in,width=2.1in]{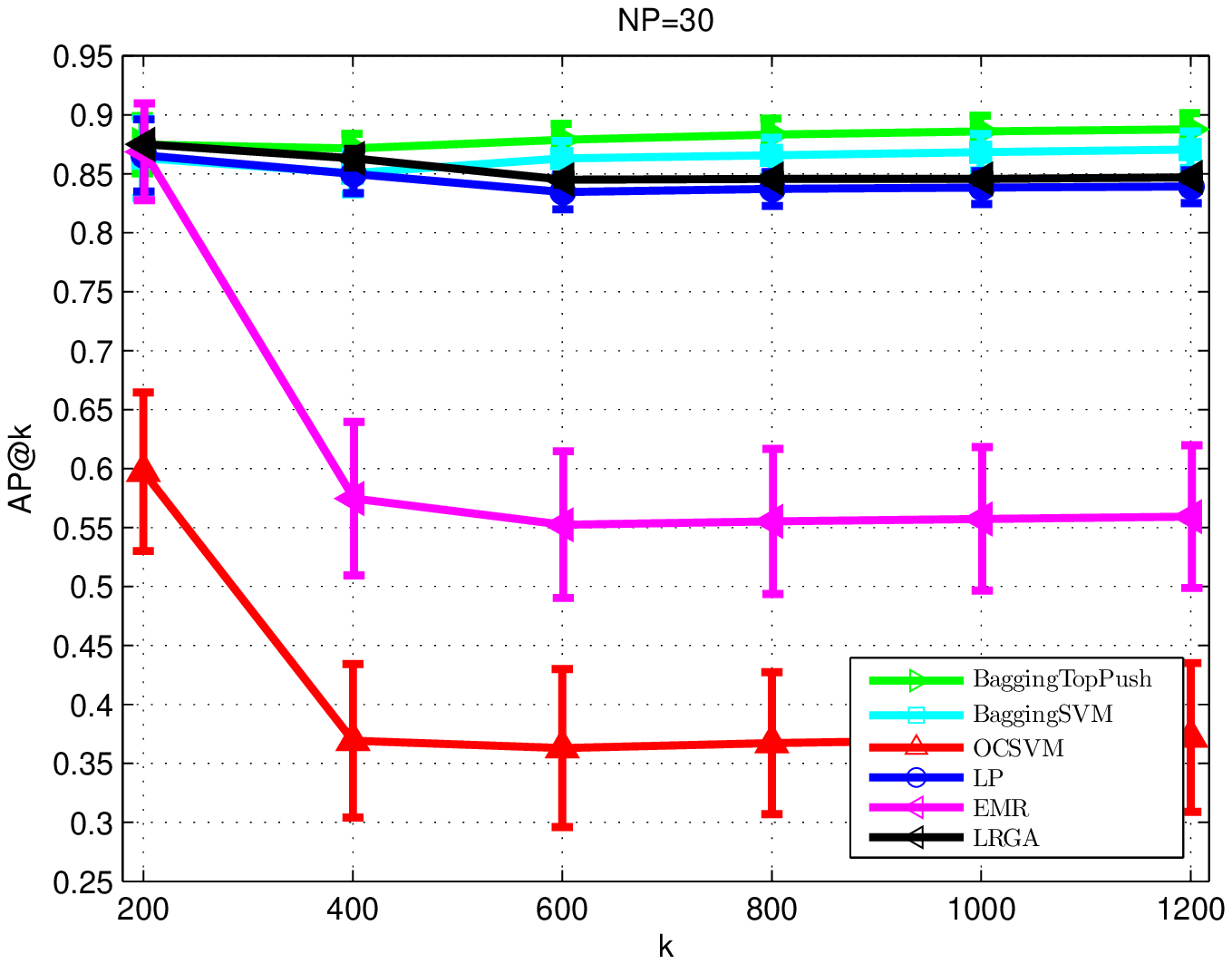}}
\caption{A comparison of average $\mathrm{AP}@k$ for different settings of $k$ and the number of positive training samples NP.}
\label{fig4}
\end{figure*}

To compare the  ranking effectiveness at the top of a ranked list, the average performance (mean$\pm$std) of different methods are shown in Fig. \ref{fig2}, \ref{fig3} and \ref{fig4}. From the results, we can find that LRGA obtains better precision and recall when NP=1(as seen in the left panels of Fig 2,3,4), and BaggingTopPush obtains better results when NP$>$1(as seen in the middle and right panels of Fig 2,3,4). This is consistent with the design of BaggingTopPush that aims to maximise the accuracy at the top of the ranked list. Compared to other methods, the performance of OCSVM is worst, and this is related to the fact that OCSVM usually needs a sufficiently large number of positive samples. Although LP and EMR have been successfully used in \citep{si2014search,Si2015carbon} to search carbon stars from SDSS and LAMOST respectively, the experimental results show that LP or EMR is not the best method to retrieval carbon stars. Especially, the performance of EMR is poor compared to BaggingTopPush or LRGA.

In terms of evaluation metric AUC, which measures the global ranking performance of the model, we list the average results in Table \ref{AUC}. We can see that BaggingTopPush achieves best AUC values for most NP values (only worse than LRGA when NP=1). This indicates that BaggingTopPush not only has better accuracy at the top of ranked list, but also has excellent global ranking performance. In rare astrophysical objects searching, if only a few of positive samples ($\leq$ 3) are given, one can utilise the LRGA method to retrieve the relevant samples, otherwise, we empirically found that BaggingTopPush is the best method used to obtain an accurate ranked list.

\begin{figure*}[!htbp]
\centering
\subfigure[NP=1] {\includegraphics[height=1.6in,width=6.0in]{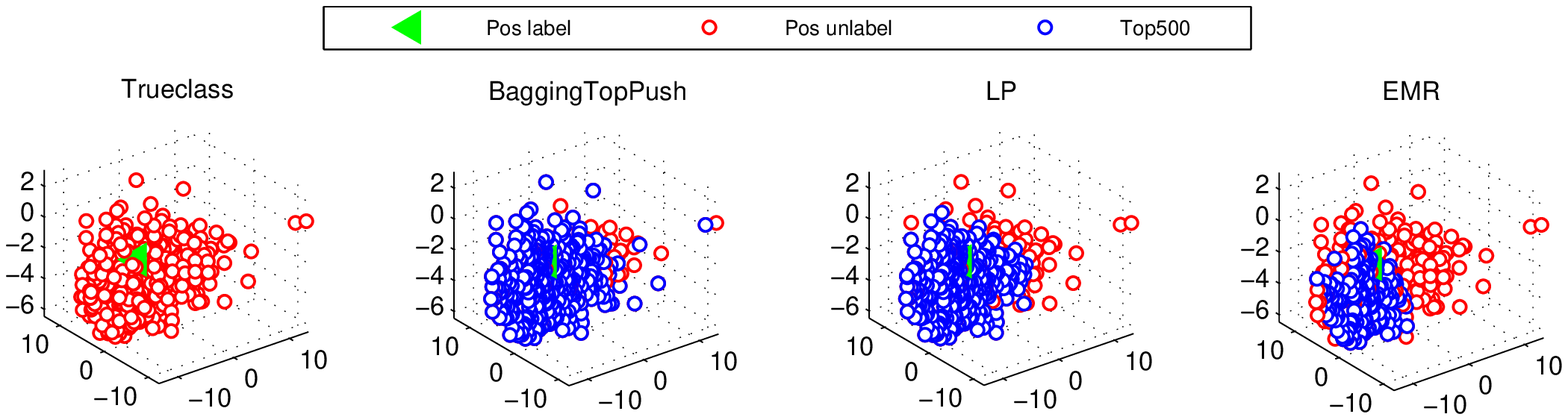}}
\subfigure[NP=10] {\includegraphics[height=1.6in,width=6.0in]{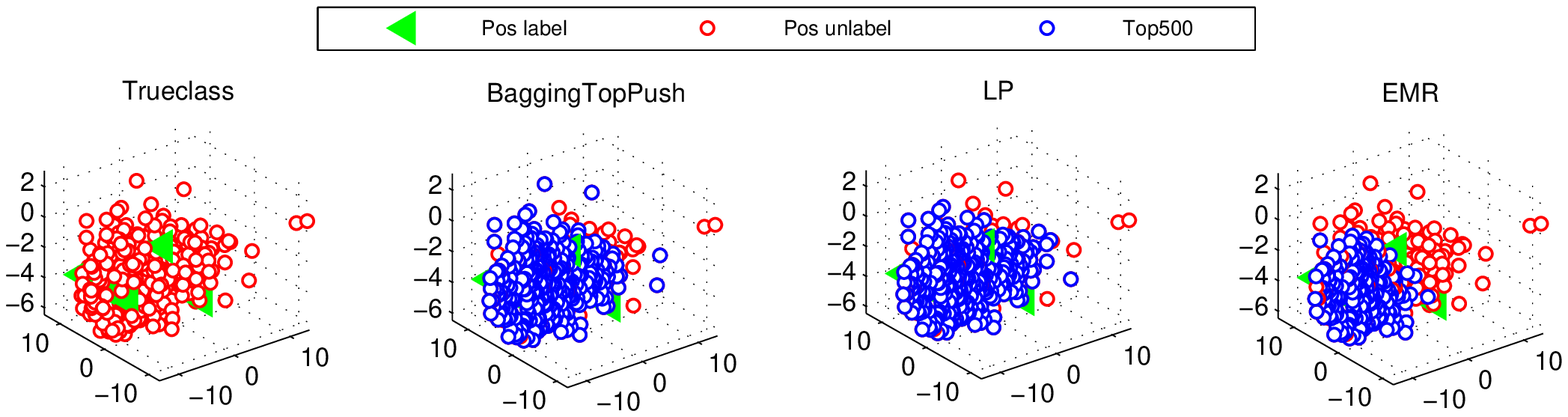}}
\subfigure[NP=30] {\includegraphics[height=1.6in,width=6.0in]{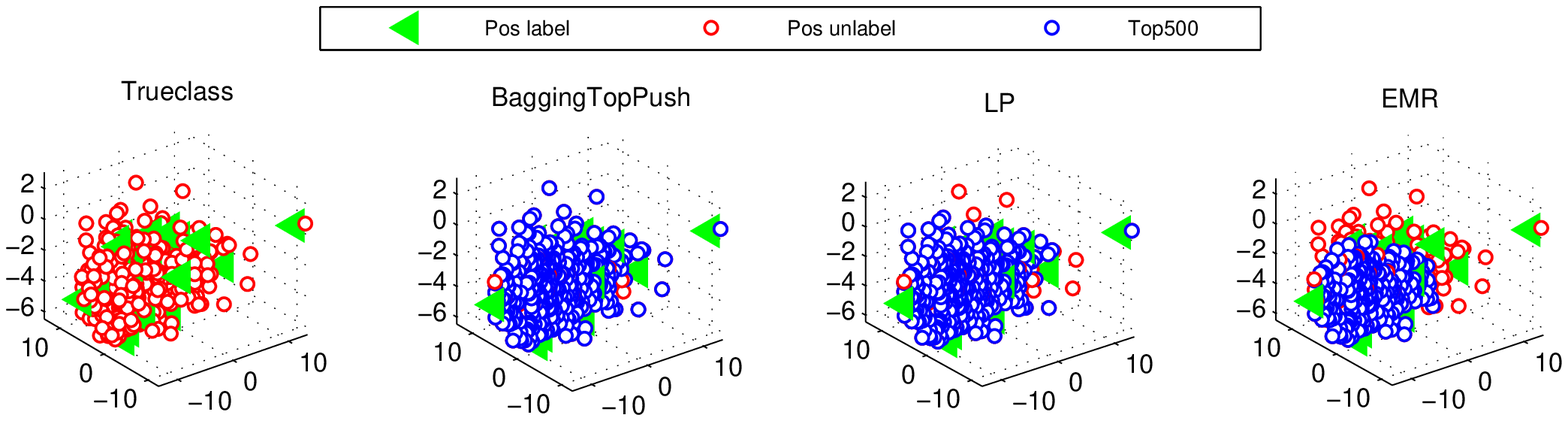}}
\caption{A comparison of carbon stars retrieval recall performance of BaggingTopPush, LP and EMR under different values of NP. The green triangles (Pos label) denote labeled carbon stars, the red circles (Pos unlabel) denote unlabeled carbon stars and the blue circles (Top500) denote retrieved carbon stars in the top 500 of ranked list.}
\label{fig5}
\end{figure*}

To intuitively compare retrieval recall performance of different methods on our spectral data set, we have visualised the retrieval results of 450 carbon stars in 3D space constructed by PCA  in Fig. \ref{fig5}. Under different numbers of positive training samples NP, it shows the data distributions of the carbon stars selected to be positive samples and the retrieved results of top 500 at the ranked list. The more blue circles, the better retrieval performance. As expected, the retrieval performances of different methods increase with NP, and BaggingTopPush always obtain better results than LP and EMR.

\begin{figure*}
\hskip -0.13in
\vskip -0.35in
\centerline{\includegraphics[height=8.0in,width=7.0in]{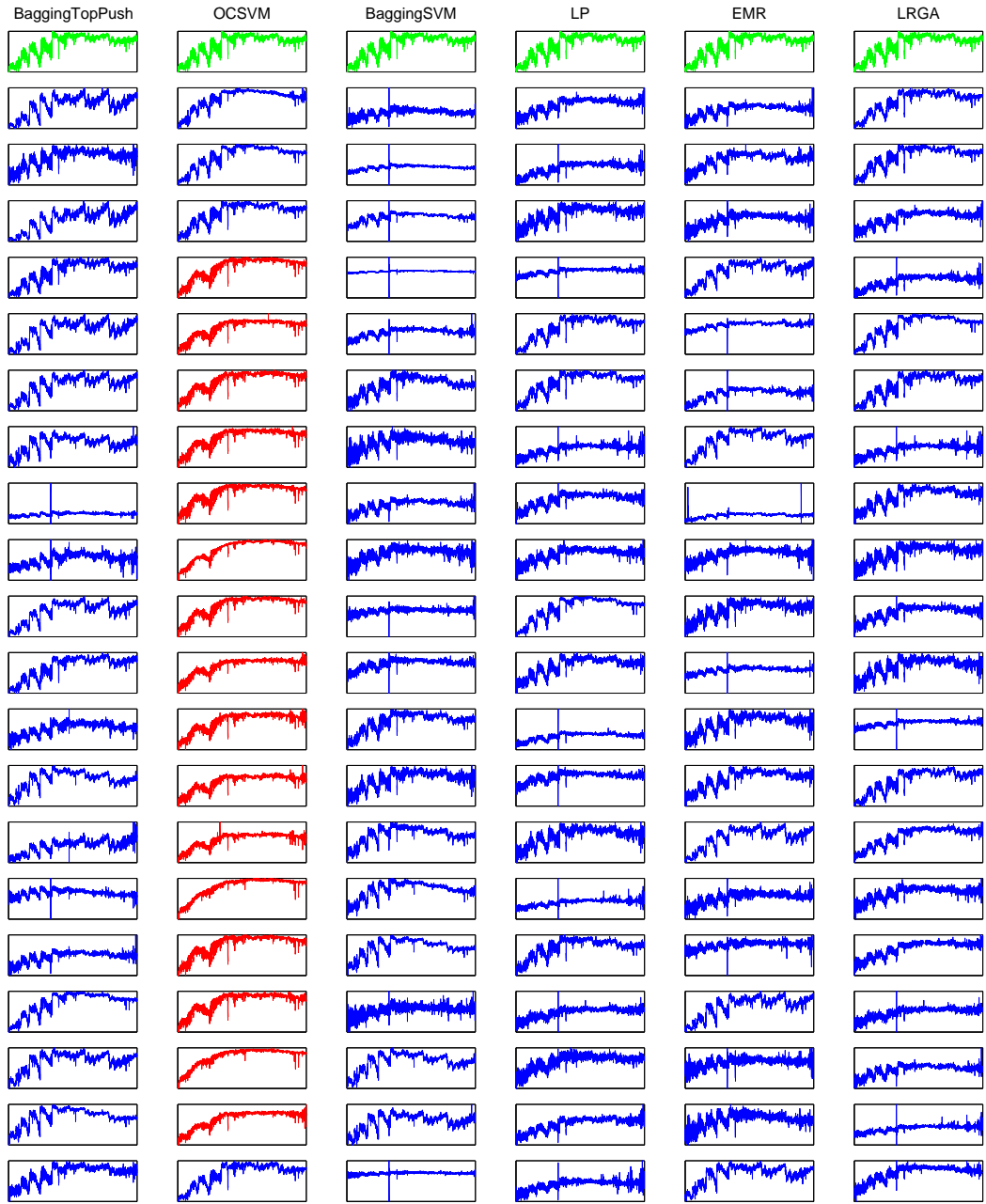}}
\vskip -0.85in
\caption{A comparison of carbon stars retrieval precision performance under NP=1. In each column, the first image with green spectra is a randomly selected positive sample (labeled carbon star), and the rest are the top 20 returns where blue color denotes it is the true carbon star and red denote it is not carbon star.}
\label{fig6}
\end{figure*}

In Fig. \ref{fig6}, we plot the top 20 spectra retrieved by each method with NP=1 to compare the correctness of different methods. From the results, we can see that the spectra retrieved by  OCSVM are mostly incorrect. Although the top 20 spectra retrieved by other 5 methods are all correct, we can further find that the results retrieved by BaggingTopPush and LRGA are visually more close to the given positive example than LP, EMR and BaggingSVM.

\subsection{Ranking efficiency}
\begin{figure}[!htbp]
\hskip -0.4in
\centerline{\includegraphics[height=1.4in,width=4.0in]{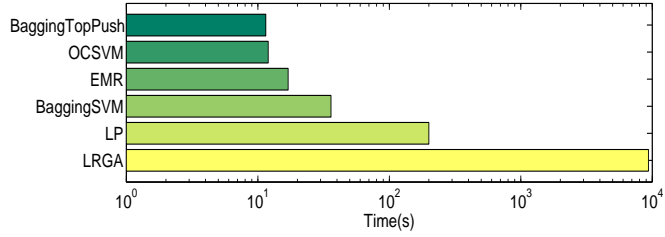}}
\caption{A comparison of CPU time of different methods on our data set.}
\label{fig7}
\end{figure}

To evaluate ranking efficiency, we run all 6 PU learning methods on our data set, and record their computation CPU times in Fig. \ref{fig7}. From the results, we can see that the BaggingTopPush is prominently faster than EMR, BaggingSVM, LP, and  LRGA, while the gaps between BaggingTopPush and OCSVM is very small. This result owns to the fact that the computational complexity of BaggingTopPush and OCSVM is linear in the number of training samples. So, with reasonable values of $K$, $T$ and $\lambda$, the BaggingTopPush method not only has better ranking performance but also spends remarkably less computational time.

\subsection{Influence of parameters}
The selection of the model parameters usually plays an important role to many machine learning methods because different settings of the model parameters may have impact on the performance of an algorithm directly. In the following, we evaluate the performance of our BaggingTopPush method with different values of the parameters.

\begin{figure*}[!htbp]
\centering
\subfigure[NP=1] {\includegraphics[height=1.8in,width=2.1in]{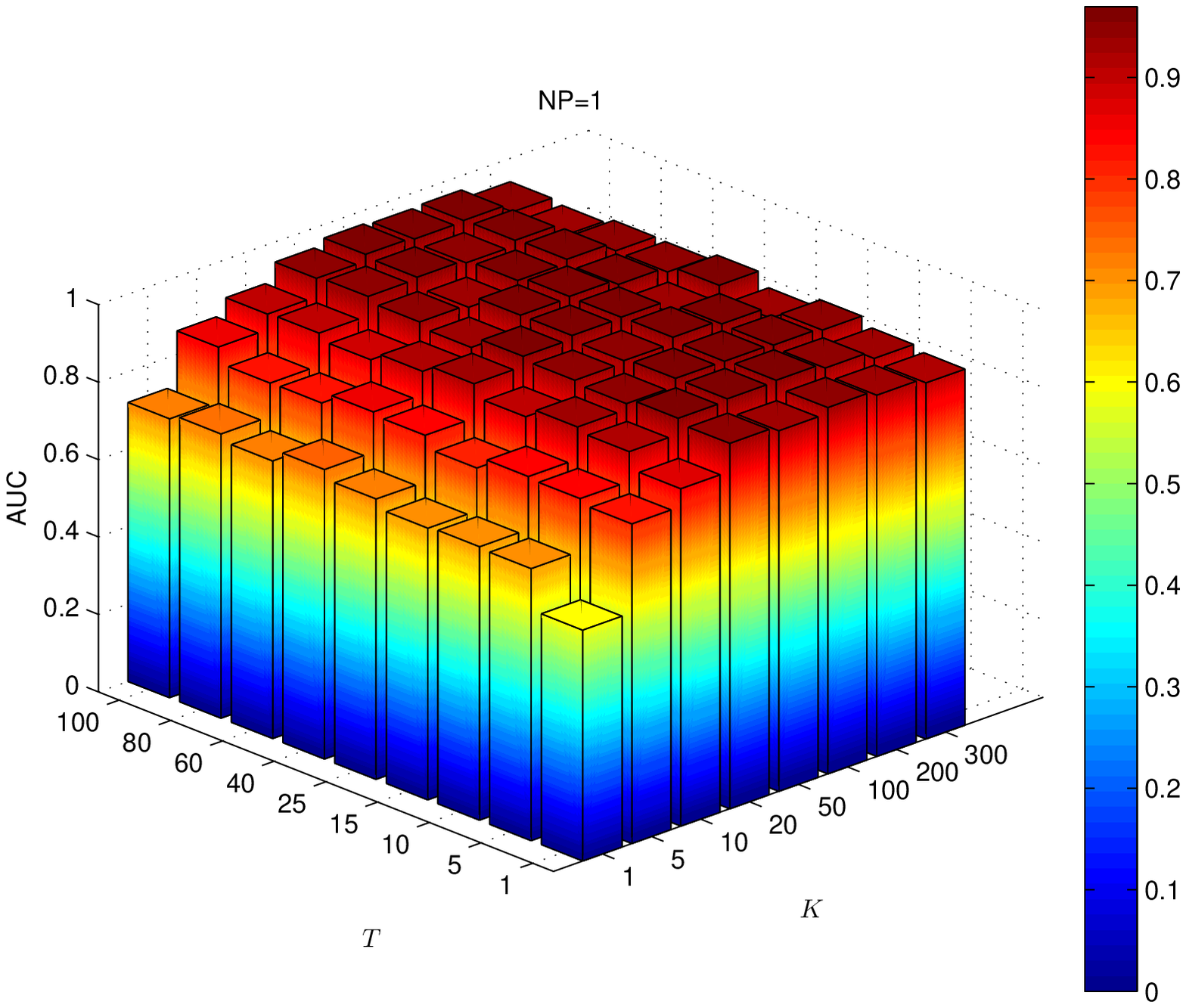}}
\subfigure[NP=10] {\includegraphics[height=1.8in,width=2.1in]{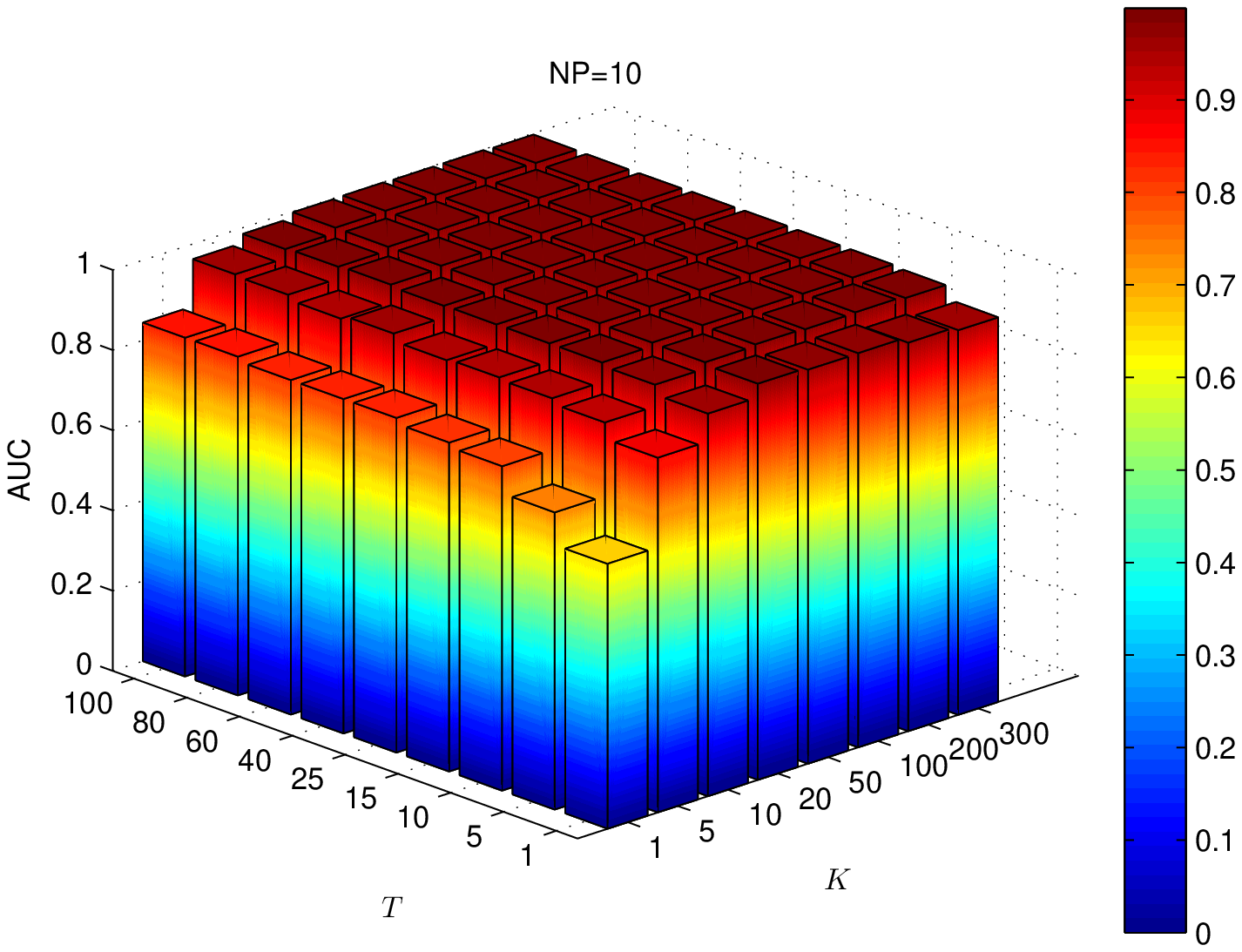}}
\subfigure[NP=30] {\includegraphics[height=1.8in,width=2.1in]{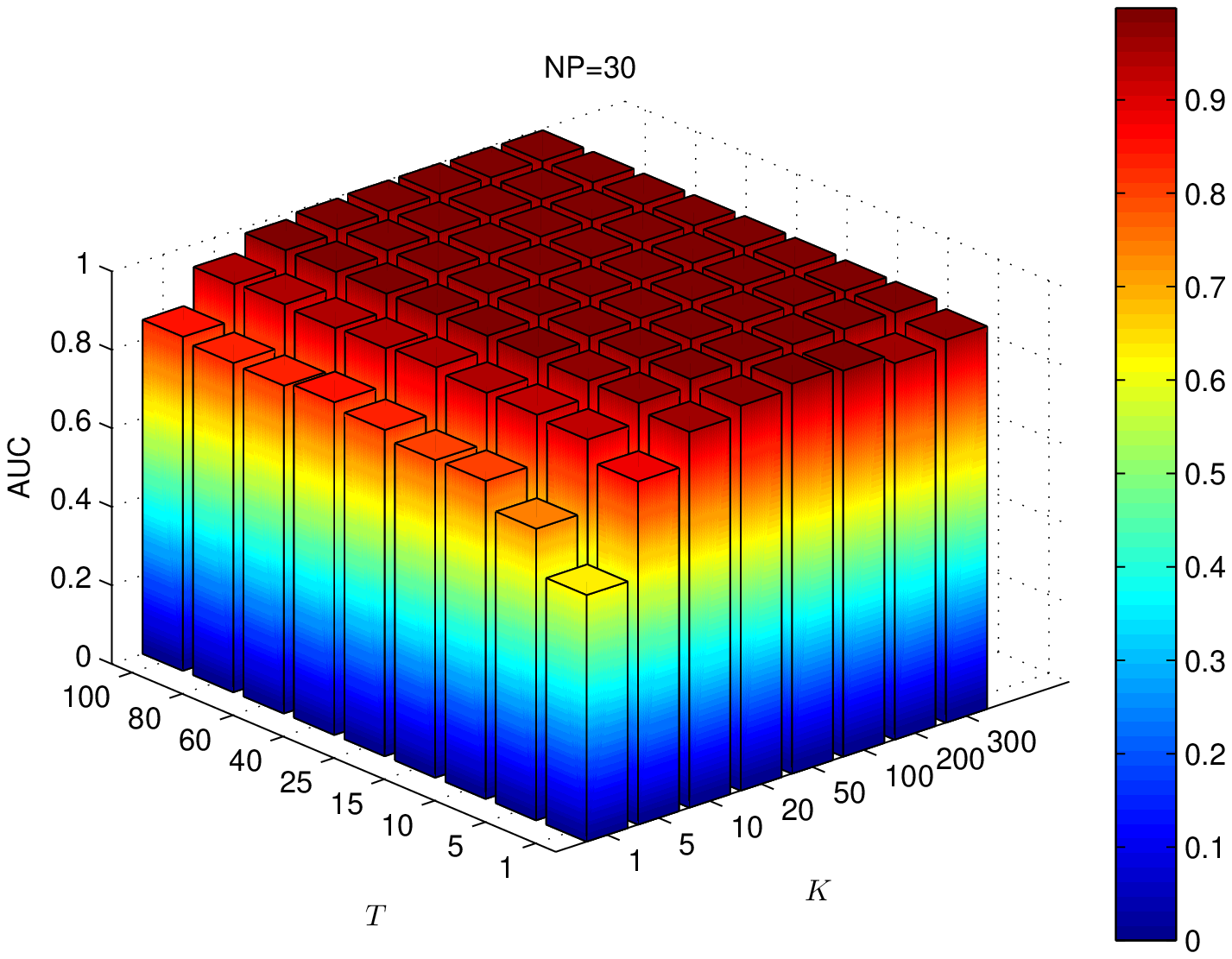}}
\caption{Performance variations of BaggingTopPush with respect to the number of bootstraps $T$ and the size of bootstrap samples $K$, for different values of NP.}
\label{fig8}
\end{figure*}

As shown in Algorithm 1, there are three parameters in BaggingTopPush method: $K$, $T$, and $\lambda$. Parameter $K$ is the number of samples randomly selected from $\mathbf{U}$ by one bootstrap and parameter $T$ is the number of bootstraps. Fig.\ref{fig8} shows the performance variations of BaggingTopPush with respect to $T$, $K$, and different values of NP.  From Fig.\ref{fig8}, the performance of BaggingTopPush is not sensitive to the selection of $K$ and $T$ when $K>50$ \& $T>40$. Intuitively, the larger $K$ and $T$, the more time cost in ranking, thus we just select  $K=200$ and $T=50$ in our experiments. In spectral retrieval applications, the selection of parameter $K$ is related to the number of positive samples and the true proportion of positive samples hidden in $\mathbf{U}$, so it is a parameter that needs to be tuned based on specific cases.

\begin{figure}[!htbp]
\centerline{\includegraphics[height=2.0in,width=2.8in]{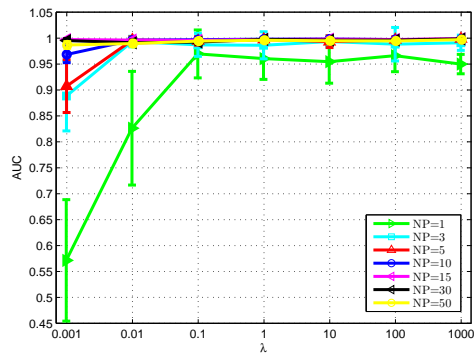}}
\caption{Performance variations of BaggingTopPush as a function of regularization parameter $\lambda$, for different values of NP.}
\label{fig9}
\end{figure}

Fig. \ref{fig9} shows the performance variations of BaggingTopPush as a function of regularisation parameter $\lambda$ for different values of NP. We can find that the ranking performance is closely related to $\lambda$ only when the number of positive training samples $\mathrm{NP}\leq5$. When $\mathrm{NP}\geq10$, the ranking performance is not sensitive to the selection of $\lambda$. Since the computational cost of TopPush reduces when a larger value of $\lambda$ is used \citep{NIPS2014_5222}, we set $\lambda=100$ in our experiments.

\section{Discussion and Conclusions}
In rare spectra retrieval application, how to extract the key features from an initial set of spectral data to facilitate the subsequent learning is a challenging problem. Since the features in carbon star spectra are very broad, we directly apply PCA to get the most informative features. However, if the spectral features of some rare type objects are sharp or indistinct, we need to extract the informative features carefully by defining some spectral line indices.

In this paper, we focus on the PU learning problems in rare astrophysical objects searching, and present the BaggingTopPush approach to retrieve the rare spectra in massive astronomical datasets. Based on the bipartite ranking model and bagging techniques, the new method aggregates bipartite ranking functions which are trained to place positive samples ahead of a small random subsample of all the unlabelled samples. The proposed method has the merit of high accuracy at the top of the ranked list, which is useful in searching for rare astronomical objects. Compared with previous algorithms used to search for rare spectra, BaggingTopPush not only has better retrieval performance but also spends remarkably less computational time. We also investigated the influence of model parameters on ranking performance, and provided optimal parameter choices making our method simple to apply.

The source code of BaggingTopPush for rare spectra retrieval is publicly available, and can be downloaded at \url{http://paperdata.china-vo.org/AstroDM/BaggingTopPush.zip}.

\acknowledgments
This work is supported by the National Natural Science Foundation of China (Grant Nos 11390371, 11233004), and the National Key Basic Research Program of China (Grant No. 2014CB845700).

\clearpage

\clearpage

\clearpage

\begin{table*}[!htbp]
\scriptsize
\centering
\setlength{\belowcaptionskip}{30pt}
\caption{The average AUC (mean$\pm$std) of different methods with the number of positive training samples NP.}
\label{AUC}
\begin{tabular}{c|c|c|c|c|c|c}
\hline
 NP                & OCSVM                       & LP                           & EMR                        & LRGA                      & BaggingSVM                & BaggingTopPush       \\ \hline\hline
 1                 & 0.689$\pm$0.033             & 0.959$\pm$0.033              & 0.809$\pm$0.036            & \textbf{0.988}$\pm$0.015  & 0.836$\pm$0.035          & 0.966$\pm$0.040  \\
 3                 & 0.750$\pm$0.028             & 0.984$\pm$0.013              & 0.884$\pm$0.021            & 0.987$\pm$0.021           & 0.954$\pm$0.019          & \textbf{0.990}$\pm$0.017  \\
 5                 & 0.790$\pm$0.023             & 0.983$\pm$0.012              & 0.893$\pm$0.021            & 0.984$\pm$0.012           & 0.975$\pm$0.014          & \textbf{0.997}$\pm$0.014  \\
 10                & 0.837$\pm$0.021             & 0.977$\pm$0.011              & 0.906$\pm$0.015            & 0.979$\pm$0.002           & 0.991$\pm$0.010          & \textbf{0.998}$\pm$0.001           \\
 15                & 0.856$\pm$0.020             & 0.975$\pm$0.006              & 0.907$\pm$0.014            & 0.973$\pm$0.003           & 0.995$\pm$0.003          & \textbf{0.998}$\pm$0.002           \\
 30                & 0.885$\pm$0.018             & 0.958$\pm$0.004              & 0.889$\pm$0.013            & 0.961$\pm$0.002           & 0.995$\pm$0.002          & \textbf{0.997}$\pm$0.003  \\
 50                & 0.899$\pm$0.016             & 0.938$\pm$0.004              & 0.875$\pm$0.009            & 0.939$\pm$0.003           & 0.988$\pm$0.006          & \textbf{0.995}$\pm$0.003   \\ \hline\hline
 Av.               & 0.816$\pm$0.023             & 0.968$\pm$0.012              & 0.880$\pm$0.019            & 0.973$\pm$0.008           & 0.962$\pm$0.013          & \textbf{0.992}$\pm$0.013  \\ \hline
\end{tabular}
\end{table*}


\begin{thebibliography}{}
\bibitem[Amini et al.(2008)]{amini2008boosting} Amini, M. R.,  Truong, T. V., \& Goutte, C., 2008, SIGIR, 99--106
\bibitem[Ahn et al.(2014)]{ahn2014tenth} Ahn, C. P., Alexandroff, R., Allende P. C.,  Anders, F., Anderson, S. F., Anderton, T., Andrews, B. H., Aubourg, {\'E}.,  Bailey, St., Bastien, F. A., \& others, 2014, ApJS, 211, 17
\bibitem[Boyd et al.(2012)]{boyd2012accuracy} Boyd, S., Cortes, C., Mohri, M., \& Radovanovic, A., 2012, NIPS, 953--961
\bibitem[Breiman(1996)]{breiman1996bagging} Breiman, L., 1996, Machine Learning, 24, 123
\bibitem[Cl{\'e}men{\c{c}}on \&  Vayatis(2007)]{clemenccon2007ranking} Cl{\'e}men{\c{c}}on, S., \&  Vayatis, N., 2007, JMLR, 8, 2671
\bibitem[Chang \& Lin(2011)]{CC01a} Chang, C. C., \& Lin, C. J., 2011, ACM TIST, 2, 1--27
\bibitem[Kotlowski et al.(2011)]{kotlowski2011bipartite} Kotlowski, W., Dembczynski, K. J., \& Huellermeier, E., 2011, ICML, 1113--1120
\bibitem[Liu et al.(2003)]{liu2003building} Liu, B., Dai, Y.,  Li, X., Lee, W. S. \&  Yu, P. S.,  2003, ICDM, 179--186
\bibitem[Liu (2009)]{liu2009learning} Liu, T.Y., 2009, Foundations and Trends in Information Retrieval, 3, 225
\bibitem[Li et al.(2014)]{NIPS2014_5222}  Li N.,  Jin R., \&   Zhou Z. H., 2014, NIPS, 1502--1510
\bibitem[Lee et al.(2013)]{61acebb7c98a4881b4c72e29e6d6ea8d} Lee, C. H., Oluwasanmi K., \& Joydeep G., 2013, EMBC, 3459-3462
\bibitem[Manevitz \& Yousef(2002)]{manevitz2002one} Manevitz, L. M., \& Yousef, M., 2002, JMLR, 2, 139
\bibitem[Mordelet \&  Vert(2014)]{mordelet2014bagging} Mordelet, F., \&  Vert, J. P., 2014, Pattern Recognition Letters, 37, 201
\bibitem[Mordelet \& Vert(2011)]{mordelet2011prodige} Mordelet, F., \& Vert, J. P., 2011, BMC Bioinformatics, 12, 389
\bibitem[Narasimhan \& Agarwal(2013)]{narasimhan2013relationship} Narasimhan, H., \& Agarwal, S., 2013, NIPS, 2913--2921
\bibitem[Rendle et al.(2009)]{rendle2009learning} Rendle, S., Balby M. L., Nanopoulos, A., \& Schmidt T. L., 2009, SIGKDD, 727--736
\bibitem[Sch{\"o}lkopf et al.(1999)]{scholkopf1999support} Sch{\"o}lkopf, B., Williamson, R. C., Smola, A. J., Shawe-Taylor, J., \& Platt, J. C.,  1999, NIPS, 12, 582--588
\bibitem[Si et al.(2014)]{si2014search} Si, J. M.,  Luo, A. L.,  Li, Y. B., Zhang, J. N., Wei, P., Wu, Y. H., Wu, F. C., \& Zhao, Y. H., 2014, Science China Physics, Mechanics and Astronomy, 57, 176--186
\bibitem[Si et al.(2015)]{Si2015carbon} Si, J. M., Li, Y. B., Luo, A. L.,  Tu L. P.,  Shi Z. X., Zhang, J. N., Wei P.,  Zhao, G.,  Wu, Y. H., Wu, F. C., Zhao, Y. H.,  \& others 2015, RAA, 15, 1671
\bibitem[Tax \& Duin(2004)]{tax2004support} Tax, D. M., \& Duin, R. P., 2004, Machine Learning, 54, 45
\bibitem[Wang \&  Zhang(2008)]{wang2008label} Wang, F., \&  Zhang, C.,  2008, TKDE, 20, 55
\bibitem[Xu et al.(2011)]{xu2011efficient} Xu, B., Bu, J.,  Chen, C. Cai, D., He, X.,  Liu, W.,  \&  Luo, J., 2011, SIGIR, 525--534
\bibitem[Xu et al.(2009)]{yang2009ranking} Yang, Y., Xu, D., Nie, F., Luo, J., \& Zhuang, Y., 2009, ACM Multimedia, 175--184
\bibitem[Zhang \& Zhou(2009)]{zhang2009non} Zhang, Y., \& Zhou, Z.H.,  2009, IJCAI, 1357--1362
\bibitem[Zhou et al.(2004)]{zhou2004learning} Zhou, D., Bousquet, O., Lal, T. N.,  Weston, J., \& Sch{\"o}lkopf, B.,  2004, NIPS, 16, 321
\bibitem[Zhou et al.(2004)]{zhou2004ranking} Zhou, D., Weston, J., Gretton, A., Bousquet, O., \& Sch{\"o}lkopf, B.,  2004, NIPS, 16, 169
\end{thebibliography}
\end{document}